\documentclass[aps,prb,a4paper,twocolumn,showpacs]{revtex4}
\usepackage{graphicx}
\usepackage{epstopdf}
\usepackage{amsmath}
\usepackage{amsfonts}
\usepackage{amssymb}
\usepackage{subfigure}
\usepackage{color}

\begin{document}
\title{Intravalley Andreev reflection in the multi-terminal device with Y-shaped Kekul\'{e} graphene superlattices}
\author{Chao Wang$^{1,}$$^{2}$,  Peipei Zhang$^{2}$, Yu-Xian Li$^{2}$$^{\ast}$, Juntao Song$^{2}$$^{\ast}$}
\affiliation{$^{1}$College of Physics, Shijiazhuang University, Shijiazhuang 050035, People's Republic of China\\
$^{2}$College of Physics $\&$ Information Engineering and Hebei Advanced Thin Films Laboratory,
Hebei Normal University, Shijiazhuang 050024, People's Republic of China
}

\begin{abstract}
Using the tight-binding model, a multi-terminal superconductor(S) device is proposed, where the structures of the center region are primitive grpahene(G) and Y-shaped Kekul\'{e} graphene superlattice(GS), respectively. The intravalley Andreev reflection is studied in this model through the utilization of the non-equilibrium Green's function method. In the G/S device, due to the time-reverse symmetry the dominant process of Andreev reflection is intervalley reflection. In a three-terminal GS/S device, it has been observed that the coefficient of intravalley Andreev reflection can surpass that of intervalley reflection in crossed Andreev reflection. This is attributed to the coupling between valleys K and -K within the first Brillouin zone, resulting in enhanced intervalley scattering.  The valley-dependent transport of Andreev reflection can be influenced by the phase difference between superconductor terminals. The valley polarization of  the local Andreev reflection and the crossed Andreev reflection could be controlled by adjusting the structure of the central region.
\end{abstract}

%\pacs{74.45.+c, 73.23.-b, 75.47.Jn}

\date{\today}

\maketitle

\vskip2pc
\narrowtext
\section{Introduction}
Since graphene(G)\cite{Novoselov,Neto,Haldane,Novoselov2005,Cheianov,Young,Nakada,Geim,Sarma} was successfully exploited in the laboratory, which is essentially a single layer of pure carbon atoms bonded together in a honeycomb lattice, it has been the most interesting two-dimensional material in the research community due to its special properties. As is known, the unit cell of graphene  consists of two structure-different carbon A and B atoms. The six corners of the hexagonal Brillouin zone are divided into two inequivalent groups, labeled as the $K$ and $-K$ valleys. The two valleys are related by the time reversal symmetry and the spatial inversion symmetry, so electrons in graphene have extra pseudo-spin and valley degrees of freedom besides the usual charge and spin ones.
\begin{figure}[t]
%\captionsetup[subfigure]{labelformat=simple}
%\centering
%\subfloat[]{\label{Fig:R1}%%
  %{\includegraphics[width=0.45\columnwidth]{fig1a.pdf}}
  \includegraphics[width=1.0\columnwidth]{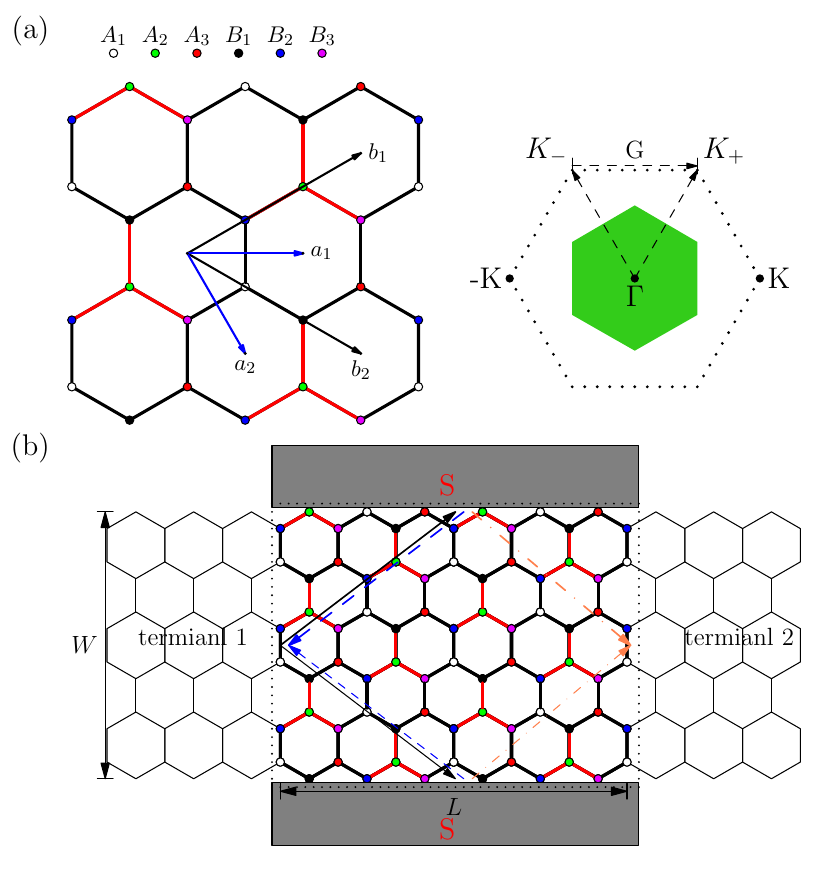}
  %{\includegraphics[width=0.5\columnwidth]{fig1b.pdf}}\\[20pt]
  %\centering
  %\hspace{0.0\linewidth}
  %{\includegraphics[width=0.5\columnwidth]{bofG.pdf}}
   \hspace{0.0\linewidth}
  %{\includegraphics[width=1.0\columnwidth]{band2.pdf}}
\caption{(Color online) (a) The diagram of the Y-shaped Kekul\'{e} graphene superlattice and the Brillouin zone(green hexagon). The two basis vectors for the graphene (superlattices) are labeled as $a_1$($b_1$) and $a_2$($b_2$). (b) The four-terminal graphene-superconductor device with the Y-shaped Kekule superlattice. The incident electrons travel into the superconductors, which are represented by the solid arrow lines (black lines), and the reflected holes travel back to the terminal $1$(blue lines) or terminal $2$ (brown lines), respectively. Terminal $1$ and $2$ are the primitive graphene ribbons. $L=2$ ($W=3$) is the length(width) of the center region of the graphene-superconductor junction. }\label{fig1}
\end{figure}

The large momentum difference between the two valleys in clean graphene samples effectively suppresses intervalley scattering, thereby conserving the valley degree of freedom as a quantum number in electron transports. This conservation can be harnessed to utilize the valley degree of freedom as an information carrier\cite{Pesin2012,xiao2007}, which is similar to how  the spin of the electron is utilized.  Recently, the valley-dependent transport of electron  has drawn much attention in graphene-like materials, where the valley of the electron can be detected and manipulated. One kind of  graphene superlattice(GS) structures can be constructed by the Kekul\'{e} distortion\cite{Giovannetti2015,wang2018,Andrade2020,Kim2016}. In the Kekul\'{e} GS, the unit cell is enlarged by $\sqrt{3}\times\sqrt{3}$ so that the points $K$ and $-K$ of the pristine graphene are folded into the point $\Gamma$ of the GS structure in the first Brillouin zone, which is shown in Fig. \ref{fig1} (a). In G/GS and GS/G/GS structures,  valley inverse effect\cite{beenakker2018} and  valley supercurrent\cite{wang2020} have been predicted and studied in theory, respectively.

In a conductor-superconductor(C/S) heterojunction, the conductance is determined by the Andreev reflection\cite{Beenakker,Titov,Xie2009,Efetov,Linder,Cheng,Xing,Rainis,Yanxia,Amet2016,Schelter,Rickhaus} when the bias is less than the superconductor gap. The process of Andreev reflection\cite{Andreev,de Jong,zhang2022} is that incident electrons from the conductor are reflected as holes back to the conductor  at the interface between the conductor and superconductor.  Due to the time-reversal symmetry, the incident electron and the reflected hole belong to the different valleys in the graphene-superconductor structure. In some conditions, the valley degrees of freedom of the electron and the hole can be manipulated\cite{Niu2007,Niu2008,Geim2014,Beenakker2007,WangYao2012}. For example, when the time reversal is broken, the electron and the hole can come from the same valley\cite{wangchao2020,zeng2021}.

In our model shown in Figs. \ref{fig1} (b), it is a four-terminal device. The central region is the Y-shaped Kekul\'{e} superlattices, which are connected with two graphene terminals and two superconductor terminals. For a three-terminal device there is only one superconductor terminal.  Assuming that incident electrons from terminal 1 travel into the center region and flow to the superconductor terminal, there are holes reflected back to the graphene terminal 1 and 2. The valley degrees of freedom of the electron and the hole can exhibit either identical or opposite characteristics due to the intervalley scattering effect. When incident electrons and reflected holes have the opposite valley degree of freedom, it is intervalley Andreev reflection. Conversely, for the intravalley Andreev reflection   electrons and  holes have the same valley degree of freedom.

\begin{figure}[b]
\centering
{\includegraphics[width=0.7\columnwidth]{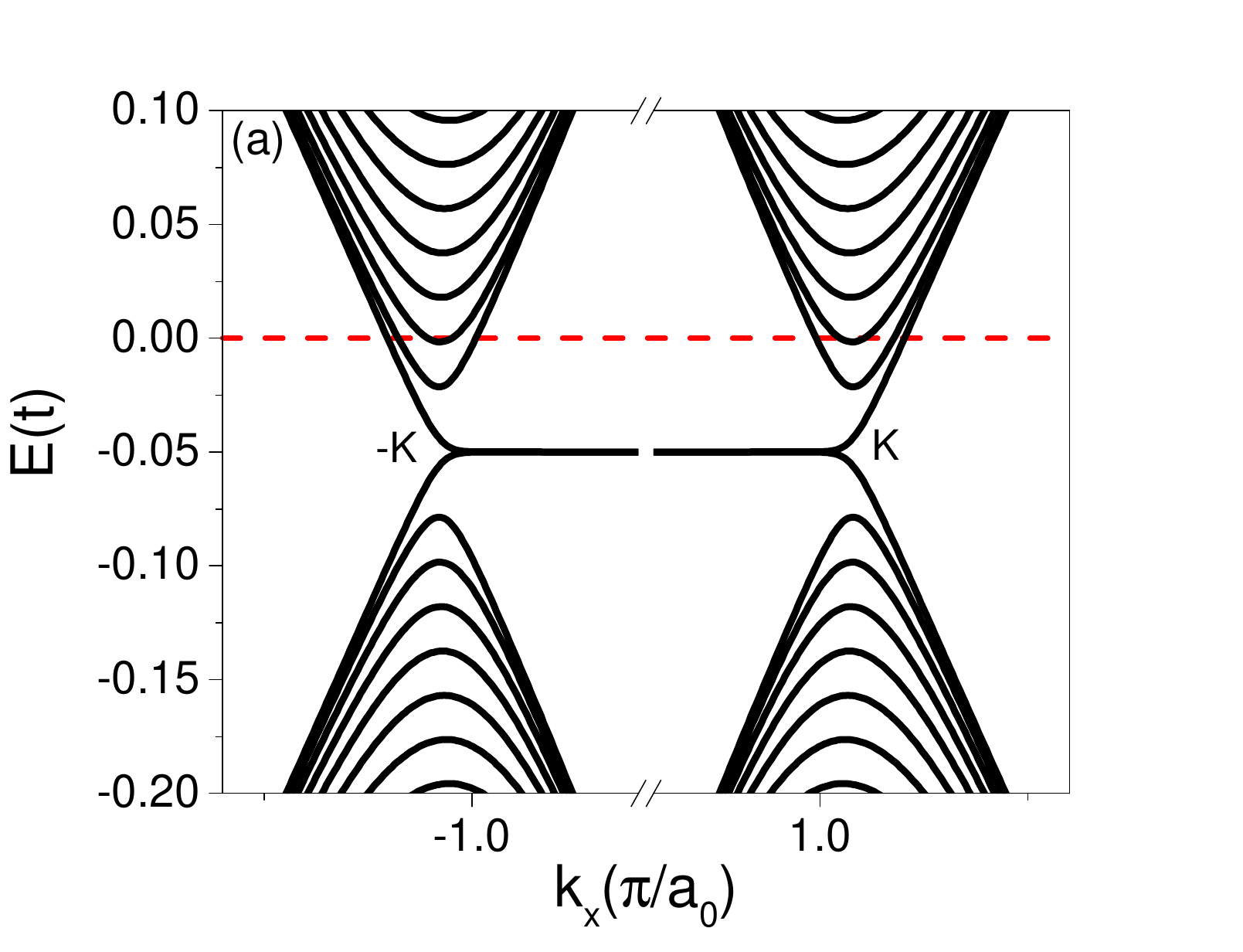}}
   \hspace{0.0\linewidth}
  {\includegraphics[width=1.0\columnwidth]{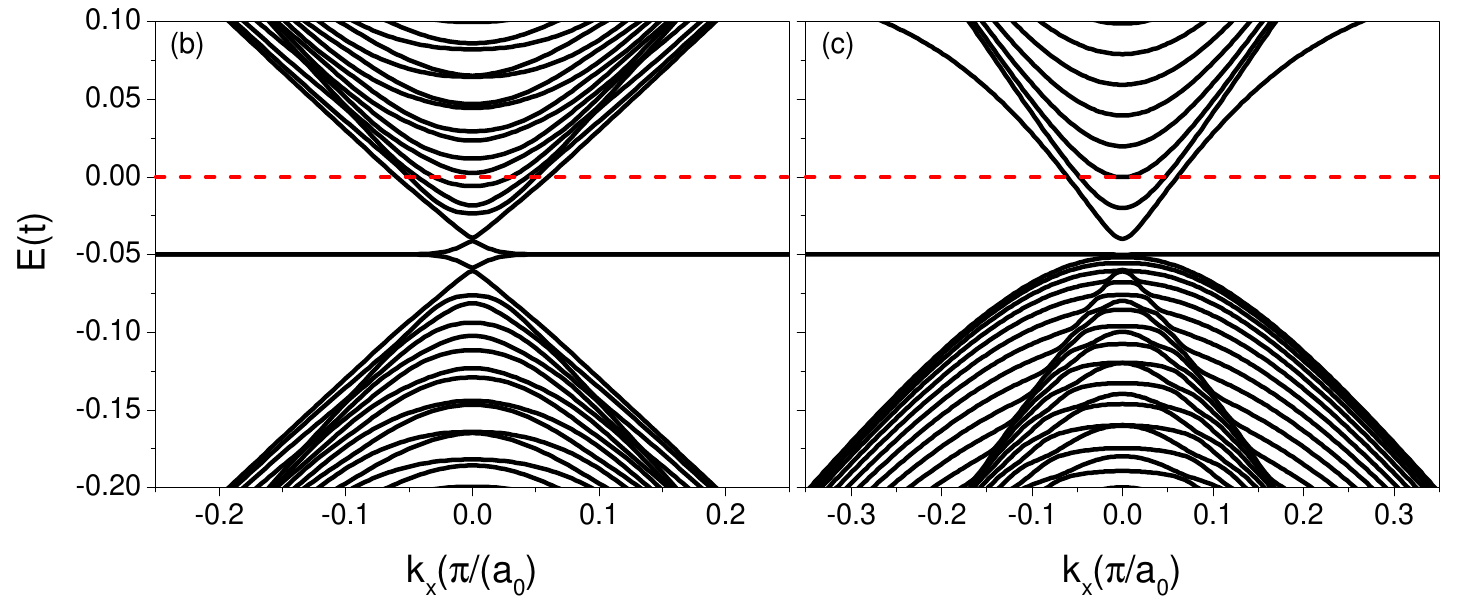}}
\caption{(Color online) The energy band for $E_0=-0.05t$. The structure are (a) Graphene ribbon, (b)  Kekul\'{e} superlattice with $U_{A_{\alpha}/B_{\alpha}}=0$  and (c) Kekul\'{e} superlattice with only $U_{A_{3}}=0.5t$. The red dashed line represents the Fermi level $E_{f}=0$. The width is $W=80$. } \label{fig2}
\end{figure}

The present paper discusses the Andreev reflection in the multi-terminal device for different valley degrees of freedom, with a particular focus on intravalley Andreev reflection. For the Y-shaped Kekul\'{e} graphene superlattice structure, whose energy band are shown in Fig. \ref{fig2} (b), the two valleys couple together, and the chirality of them are unchanged. After adjusting the on-site energy of one of six atoms in the unit cell of the superlattice, the chirality of one valley is broken and the one of the other valley is unbroken, which is showed in Fig. \ref{fig2} (c).  The coefficients of Andreev reflection, which include the local Andreev reflection and the crossed Andreev reflection represented as $T_{A11}$ and $T_{A12}$, are calculated in our works. Comparing the coefficients of Andreev reflection in the G/S structure and the GS/S structure, it is found that the intravalley Andreev reflection can be enhanced in the GS/S structure due to the intervalley scattering effect. The valley polarization of the Andreev reflection can be controlled  by adjusting the incident energy of  electrons in a three-terminal or a four-terminal model. As a consequence, the results we received provide a new way to the design and development of electronic components.

The rest of this paper is arranged as follows. In Sec.\uppercase\expandafter {\romannumeral 2}, the model Hamiltonian for the system is presented and the formalisms for calculating  the Andreev reflection coefficients and the valley polarization are derived. Our main results are shown and discussed in Sec.\uppercase \expandafter {\romannumeral 3}. Finally, a brief conclusion is presented in Sec.\uppercase\expandafter {\romannumeral 4}.\\

\section{Model and Formula}

The superconductor heterojunction investigated here is a Y-shaped Kekul\'{e} graphene superlattice in the central region, which connects with the graphene terminals and the superconductor terminals.  The Hamiltonian of the junction is
\begin{equation}\
 H=H_{GS}+H_{S}+H_{T},
\end{equation}
where $H_{GS}$, $H_{S}$ and $H_{T}$ are the Hamiltonians of the graphene superlattices ribbon, superconductor terminals, and  coupling of the center region and the superconductor terminals, respectively.

In the tight-binding representation, $H_{GS}$ is given by\cite{Neto}
\begin{equation}
\begin{aligned}
H_{GS}=
&\sum_{m,n}\sum_{\alpha=1,2,3}[E_{A_{\alpha}}a^{\dag}_{\emph{m,n}}a_{\emph{m,n}}+E_{B_{\alpha}}b^{\dag}_{\emph{m,n}}b_{\emph{m,n}}]\\
&-t_{\alpha}[a_{\emph{m,n}}^{\dagger}b_{\emph{m,n}}+a_{\emph{m,n}}^{\dagger}b_{\emph{m-1,n}}+a_{\emph{m,n}}^{\dagger}b_{\emph{m,n-1}}+H.c.],
\end{aligned}
\end{equation}
where $a_{\emph{m,n}}^{\dagger}$ ($a_{\emph{m,n}}$) and $b_{\emph{m,n}}^{\dagger}$ ($b_{\emph{m,n}}$) are the creation
(annihilation) operator  of the sublattices $A_{\alpha}$ and $B_{\alpha}$. $E_{A_{\alpha}}$  and $E_{B_{\alpha}}$ stand for the  energy of the superlattice site, where $E_{A_{\alpha}/B_{\alpha}}=E_0+U_{A_{\alpha}/B_{\alpha}}$ and $E_0$ is the primitive on-site energy and $U_{A_{\alpha}/B_{\alpha}}$ is the  modification of the on-site energy.  The second term in the Hamiltonian stands for the nearest-neighbor hopping, where $t_{\alpha}$ is the hopping energy. For the Y-shaped Kekul\'{e} superlattice the hopping energy is restructured as $t_{1/2}=t(1-\delta)$ and $t_3=t(1+2\delta)$, where $t$ is the hopping energy in the primitive graphene.  It is  considered that the region of the graphene superlattice  is directly coupled to the superconductor terminals. Described by a continuum model the superconductor terminal is represented by BCS
Hamiltonian,
\begin{equation}
H_{\emph{S}}=\sum_{\textbf{\emph{k}},\sigma}\varepsilon_{\textbf{\emph{k}}}C^{\dagger}_{\emph{\textbf{k}}\sigma}C_{\textbf{\emph{k}}\sigma}
+\sum_{\textbf{\emph{k}}}(\Delta_{0}^{}C_{\textbf{\emph{k}}\downarrow}C_{-\textbf{\emph{k}}\uparrow}
+\Delta_{0}^{*}C^{\dagger}_{\textbf{\emph{-k}}\uparrow}C^{\dagger}_{\textbf{\emph{k}}\downarrow}),
\end{equation}
where $\Delta_{0}=\Delta e^{\emph{i}\theta}$. Here $\Delta$ is the superconductor gap and $\theta$ is the
superconductor phase. The coupling between the superconductor terminal
and the center region is described by
\begin{equation}
H_{\emph{T}}=\sum_{\emph{m,n},\sigma}\emph{t}a^{\dagger}_{\emph{m,n} \sigma}C_{\sigma}(x_{\emph{i}})+h.c..
\end{equation}
Here $x_{\emph{i}}$ is the  position of the carbon atom $\emph{i}$ coupling to the superconductor terminal and
$C_{\sigma}(x)=\sum_{\emph{k}_{x}}e^{\emph{i}k_{x}x}C_{\textbf{\emph{k}}\sigma}$.

The process of the Andreev reflection is that an incident electron with the energy $\varepsilon$ flows from the graphene terminal into the superlattice region, then it is reflected as a hole at the interface between the center region and the superconductor. The reflected hole can flow out of the center region to either of the two graphene terminals. Using the nonequilibrium Green function method,  we can calculate the retarded and advanced Green function $G^{r}(\varepsilon)=[G^{a}]^{\dagger}=1/(\varepsilon I-H_{C}-\Sigma^{r})$, where $H_{C}$ is the Hamiltonian of the center region in the Nambu representation and $I$ is the unit matrix with the same dimension as $H_{C}$. The center region is the rectangular region surrounded by the dashed line in Fig. \ref{fig1} (b). $\Sigma^{r}$ are the retarded self-energy due to the coupling to the terminal $\alpha$ and the superconductor terminals\cite{Xie2009,Cheng,wangchao2020}.

\begin{figure}[tbp]
\centering
\includegraphics[width=1.0\columnwidth,angle=0]{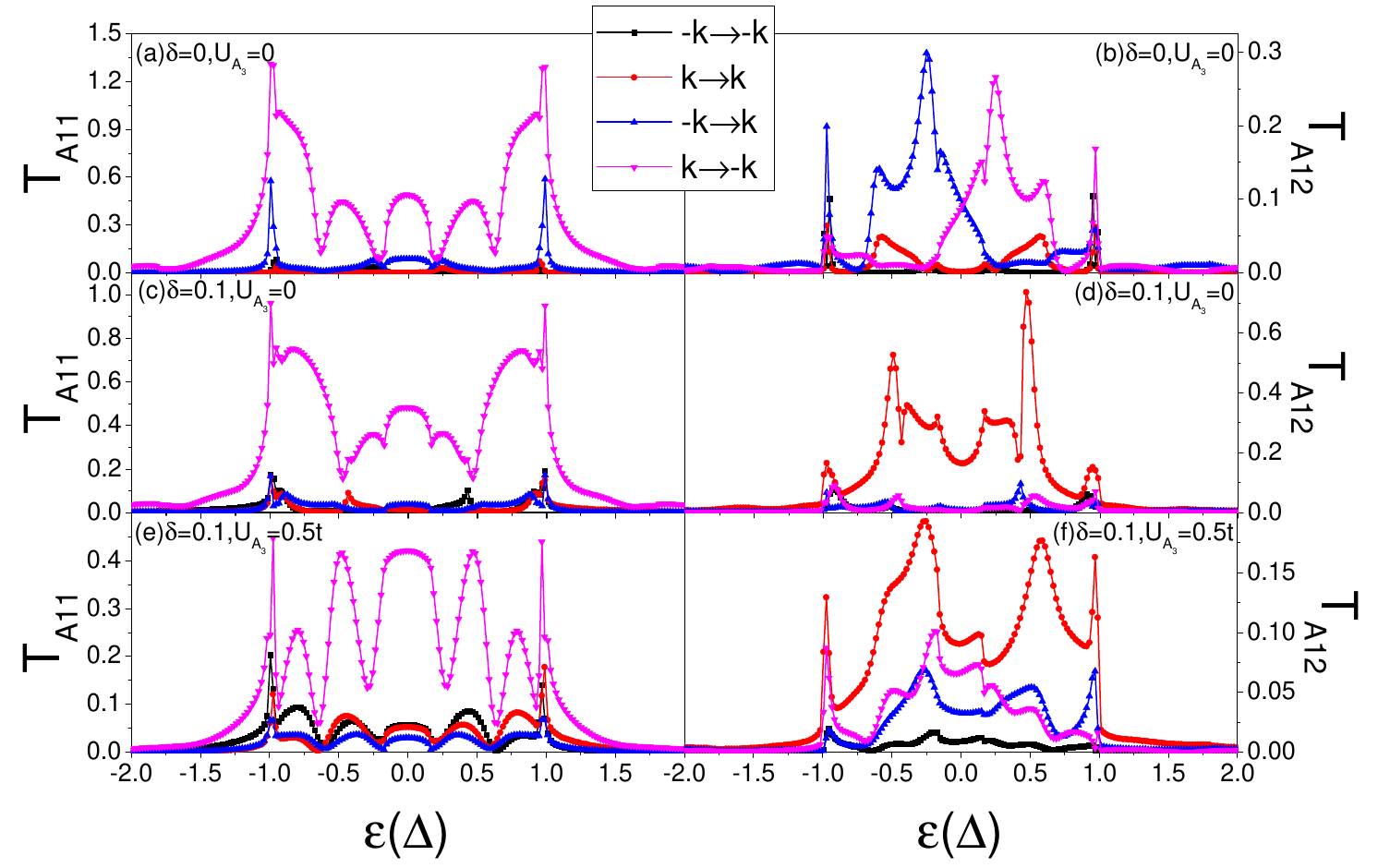}
\caption{(Color online) In three-terminal GS/S junction $T_{A,11}$ and $T_{A,12}$ vs the incident energy $\varepsilon$ with $E_{0}=-5\Delta$ when $W=80$ and $L=50$.  The structure of the center region are primitive graphene in (a) and (b), Y-shaped Kekul\'{e} superlattices with $U_{A_3}=0$ in (c) and (d),  and  superlattices with $U_{A_3}=0.5t$ in (e) and (f).}\label{fig3}
\end{figure}

In the Nambu representation, the retard Green function can be described by
\begin{equation}
{G^r} = \left( {\begin{array}{*{20}{c}}
{{G^r_{ \uparrow  \uparrow }}}&{{G^r_{ \uparrow  \downarrow }}}\\
{{G^r_{ \downarrow  \uparrow }}}&{{G^r_{ \downarrow  \downarrow }}}
\end{array}} \right),
\end{equation}
where the subscripts $\upuparrows$, $\upuparrows$,$\uparrow\downarrow$ and $\downarrow\uparrow$ represent the 11, 22, 12 and 21 matrix elements, respectively. Marking the  valley degree of freedom of the incident electron and the reflected hole as $\alpha$ and $\beta$ respectively, the  valley-dependent transport of Andreev reflection is that the incident electron with  velocity $\nu^{\alpha}_j$  from terminal 1  is reflected back as the hole with velocity $\mu^{\beta}_i$ to terminal 1 or 2, namely the local Andreev reflection or the crossed Andreev reflection. We obtain the Andreev reflection coefficients\cite{Ando1991,An2020}:
\begin{equation}
\begin{aligned}
T^{\alpha,\beta}_{A11}(\varepsilon)=\sum_{ij}\frac{\mu^{\beta}_i}{\nu^{\alpha}_j} |\Gamma^{11}_{ij}|^2,\\
T^{\alpha,\beta}_{A12}(\varepsilon)=\sum_{ij}\frac{\mu^{\beta}_i}{\nu^{\alpha}_j} |\Gamma^{12}_{ij}|^2,\\
\end{aligned}
\end{equation}
where
\begin{equation}
\begin{aligned}
\Gamma^{11}=\phi^{-1}_{h}(-)G^{r}_{\downarrow\uparrow 11}V_{e}\phi_{e}(+),\\
\Gamma^{12}=\phi^{-1}_{h}(+)G^{r}_{\downarrow\uparrow L1}V_{e}\phi_{e}(+).\\
\end{aligned}
\end{equation}
In the equations,  $\alpha(\beta)$ represents the  valley degree of freedom, which is $K$ or $-K$.  $\phi_{e}(\pm)$ and $\phi_{h}(\pm)$ are the wave function of electron and hole in the graphene terminals, respectively. $+$($-$) represents that the direction of the wave function is forward(backward).  $\chi_{e}(\pm)$ and $\chi_{h}(\pm)$ are the matrixes of the eigenvalue of the wave functions, so here is the expression $V_{e}=tI(\phi_{e}(+)\chi^{-1}_{e}(+)\phi^{-1}_{e}(+)-\phi_{e}(-)\chi^{-1}_{e}(-)\phi^{-1}_{e}(-))$. From the equations above, the polarization of the Andreev reflection is defined as
\begin{equation}
\begin{aligned}
P_{A11}=\frac{T^{K,K}_{A11}+T^{-K,K}_{A11}-T^{K,-K}_{A11}-T^{-K,-K}_{A11}}{\sum_{\alpha,\beta}T^{\alpha,\beta}_{A11}},\\
P_{A12}=\frac{T^{K,K}_{A12}+T^{-K,K}_{A12}-T^{K,-K}_{A12}-T^{-K,-K}_{A12}}{\sum_{\alpha,\beta}T^{\alpha,\beta}_{A12}}.
\end{aligned}
\end{equation}

\section{Results and discussion}
In the numerical calculation, we set the hopping energy $t=2.75$ $eV$. The length of the nearest-neighbor C-C bond is set to be $a_{0}=0.142$ $\mathrm{nm}$ as in the real graphene sample. The superconductor gap is set to be $\Delta=0.01t$. For the structure of the central region, there are three different cases. When $\delta=0$,  the center region is primitive graphene. When $\delta=0.1$, the center region is the Y-shaped Kekul\'{e} superlattice. For the superlattice structure, there are two scenarios: one where all modifications of the on-site energy are set to zero, and another where only one modification of the on-site energy is non-zero, such as $U_{A_{3}}=0.5t$. To facilitate the discussion on valley polarization, we set the Fermi energy at $E_{f}=0$.

In Fig. \ref{fig2} (a), we observe that incident electrons can originate from two distinct valleys, exhibiting different degrees of freedom in terms of valley states.  Due to $|E_{0}|=0.05t > \Delta$, the process of the Andreev reflection in our study is an  intraband reflection\cite{Beenakker,Cheng,wangchao2020}. Due to the time-reversal symmetry, in the primitive graphene, electrons and holes taking parting into the Andreev reflection  have the opposite valley degree of freedom, which is intervalley Andreev refleciton. In Figs. \ref{fig2} (b) and (c), the energy band structures of the superlattices are presented, where the two valleys are coupled together. In the superlattices, due to the coupling valleys and the stronger intervalley scattering effect,  it is possible that electrons and holes  have the same valley degree of freedom, which is intravalley Andreev reflection.

Figure \ref{fig3}  depicts  the coefficients of Andreev reflection in the three-terminal device  at  $E_{0}=-5\Delta$.   When the structure of the central region is primitive graphene, it is clear that  electrons taking part in local Andreev reflection mainly come from the valley $K$, and most of them are reflected as holes belonging to the valley $-K$, which showed in Fig. \ref{fig3} (a). For $T_{A11}$, the values of the coefficients of the intravalley Andreev reflection are always tiny. In other words, in the local Andreev reflection the intravalley Andreev refleciton is suppressed when the structure of the central region is primitive graphene, and  the dominant contribution comes from the coefficients of intervalley Andreev reflection.  The maximum value of $T^{K,-K}_{A11}$ reaches its peak at $1.35$ in Fig. \ref{fig3} (a).  It reduces to $0.9$ in Fig. \ref{fig3} (c), where the structure is superlattices. When the  modification $U_{A_{3}}=0.5t$ is introduced into the superlattices, the peak value of $T^{K,-K}_{A11}$ in Fig. \ref{fig3} (e) undergoes a reduction to $0.45$. The decrease in $T^{K,-K}_{A11}$ can be attributed to the heightened impact of intervalley scattering, which is significantly more pronounced in the superlattice configuration than in primitive graphene. As depicted in Fig. \ref{fig3} (e), a comparison with Figs. \ref{fig3} (a) and (c) reveals an augmentation of intravalley Andreev reflection.

\begin{figure}[tbp]
\centering
\includegraphics[width=1.0\columnwidth,angle=0]{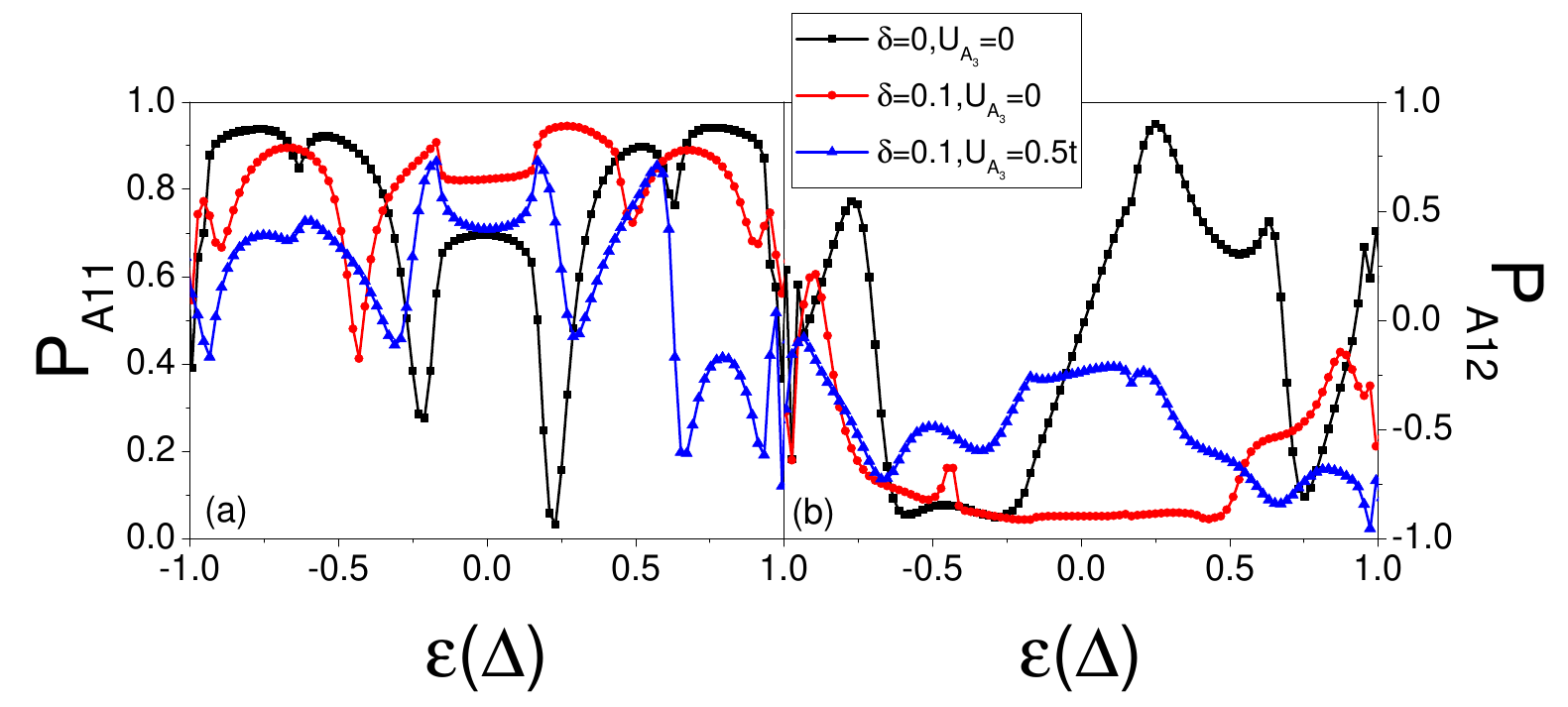}
\caption{(Color online) In the three-terminal device the valley polarization of the Andreev reflection vs $\varepsilon$ when $E_{0}=-5\Delta$  for three different center regions.}\label{fig4}
\end{figure}

For $T_{A12}$, in the primitive graphene, the intervalley reflection is the dominant process, and the intravalley reflection is nearly suppressed, which are shown in Fig. \ref{fig3} (b). When the central region adopts a Y-shaped Kekul\'{e} superlattice structure, with $\varepsilon$ varying within the range of $(-\Delta,\Delta)$, the curve value $T^{K,K}_{A12}$ consistently surpasses that of the other three curves depicted in Fig. \ref{fig3} (d). This implies that a significant portion of incident electrons originating from valley $K$ are reflected as holes back into valley $K$, ultimately flowing towards terminal 2. It explains why the peaks of $T^{K,-K}_{A11}$ reduce clearly in the Figs. \ref{fig3} (c) and (e).  In Fig. \ref{fig2} (b), we can see that the chirality of the two coupling valley is unbroken although the two valleys couple together. Due to the coupling valleys, there are  more intense intervalley scattering in the central region. In Fig. \ref{fig3} (d), the intravalley reflection in the crossed Andreev reflection is significantly enhanced due to intervalley scattering, while the intervalley reflection is effectively suppressed, exhibiting a distinct deviation from the curves observed in Fig. \ref{fig3} (b).   After introducing $U_{A_{3}}$ the chirality of one valley is broken, which is showed in Fig. \ref{fig2} (c). The breaking of chiral symmetry results in an intensified intervalley scattering effect. Therefore, in Fig. \ref{fig3} (f), although the primary contribution to $T_{A12}$ originates from $T^{K,K}_{A12}$, it is crucial not to overlook the contribution of intervalley reflections.

\begin{figure}[tpb]
\centering
\includegraphics[width=1.0\columnwidth,angle=0]{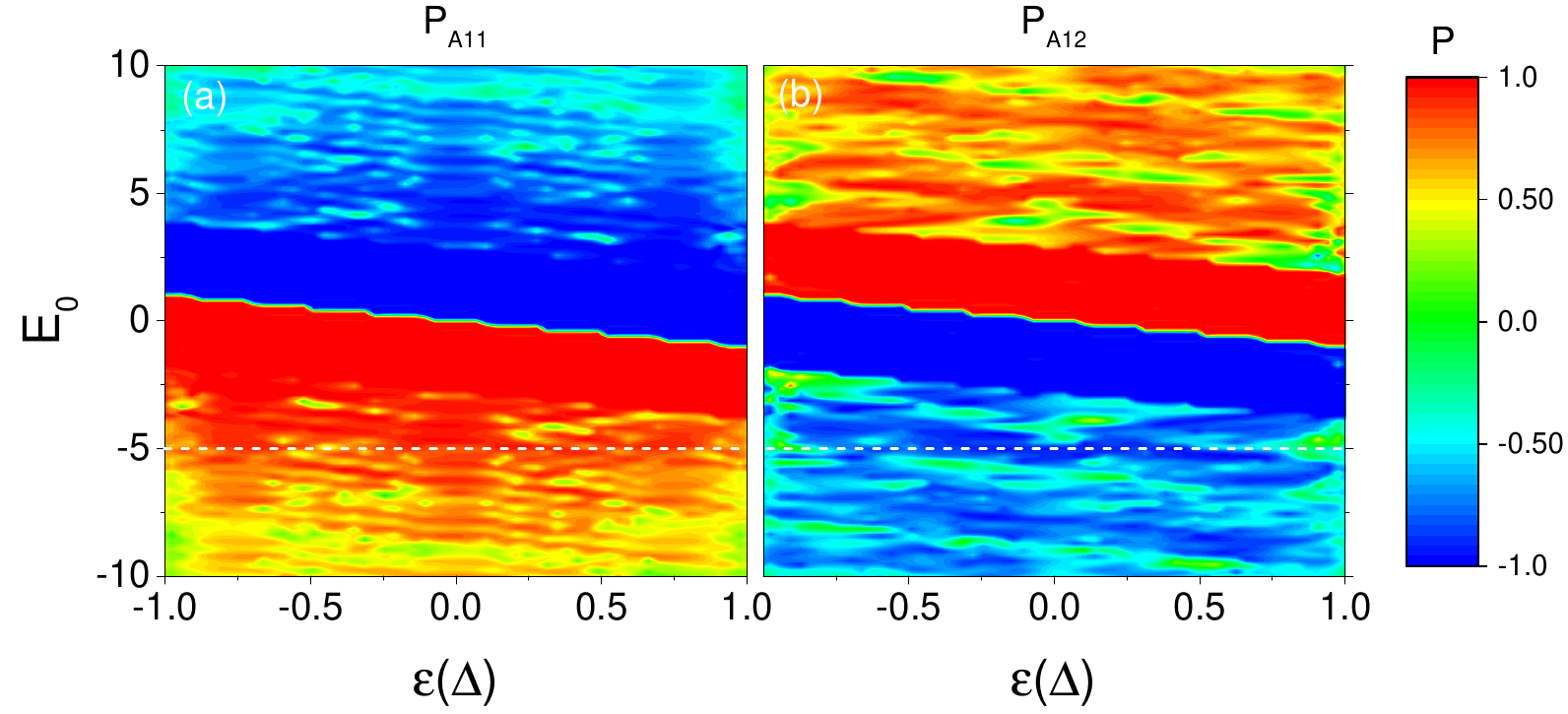}
\caption{(Color online) The valley polarization of the Andreev reflection in the three-terminal device for Y-shaped Kek superlattices with $U_{A_3}=0$.  The dashed lines corresponds to the value $E_{0}=-5\Delta$.} \label{fig5}
\end{figure}

According to the coefficients of valley-dependent Andreev reflection mentioned above, the valley polarizations of three distinct structures in the central region are calculated and depicted in Fig. \ref{fig4}. For $P_{A11}$, small plateaus can be observed around $\varepsilon=0$, while as $|\varepsilon|$ increases, the curves exhibit oscillations with consistently positive polarization values. From Fig. \ref{fig4}(a), it is evident that when transitioning from primitive graphene to a Y-shaped Kekul\'{e} superlattice structure in the central region, $P_{A11}$ increases from $0.7$ to $0.85$ at $\varepsilon=0$. However, upon introducing $U_{A_{3}}=0.5t$, $P_{A11}$ decreases back to $0.7$ again. This indicates that modulation of the center region's structure can effectively control the valley polarization of Andreev reflection.

The curves in Fig. \ref{fig4} (b) are the valley polarization of the crossed Andreev reflection with $\varepsilon$ changing. When the central region consists of primitive graphene, it exhibits a distinct behavior compared to $P_{A11}$, as $P_{A12}$ can transition from a negative to positive value within the range of $\varepsilon$ changing from $-0.5\Delta$ to $0.5\Delta$. In the case of crossed Andreev reflection in primitive graphene, the sign of valley polarization is dependent on the incident energy. By modifying the structure into superlattices, $P_{A12}$ maintains its negative value regardless of $\varepsilon$, with values reaching as low as $-0.9$ for $|\varepsilon|<0.5\Delta$. Introducing $U_{A_3}$ into these superlattices results in an increase in $P_{A12}$ to approximately $-0.25$ at $\varepsilon=0$, while still retaining its negative nature throughout. This suggests that manipulating the structure of the central region may offer potential control over valley polarization in Andreev reflection.

\begin{figure}[btp]
\includegraphics[width=1.0\columnwidth,angle=0]{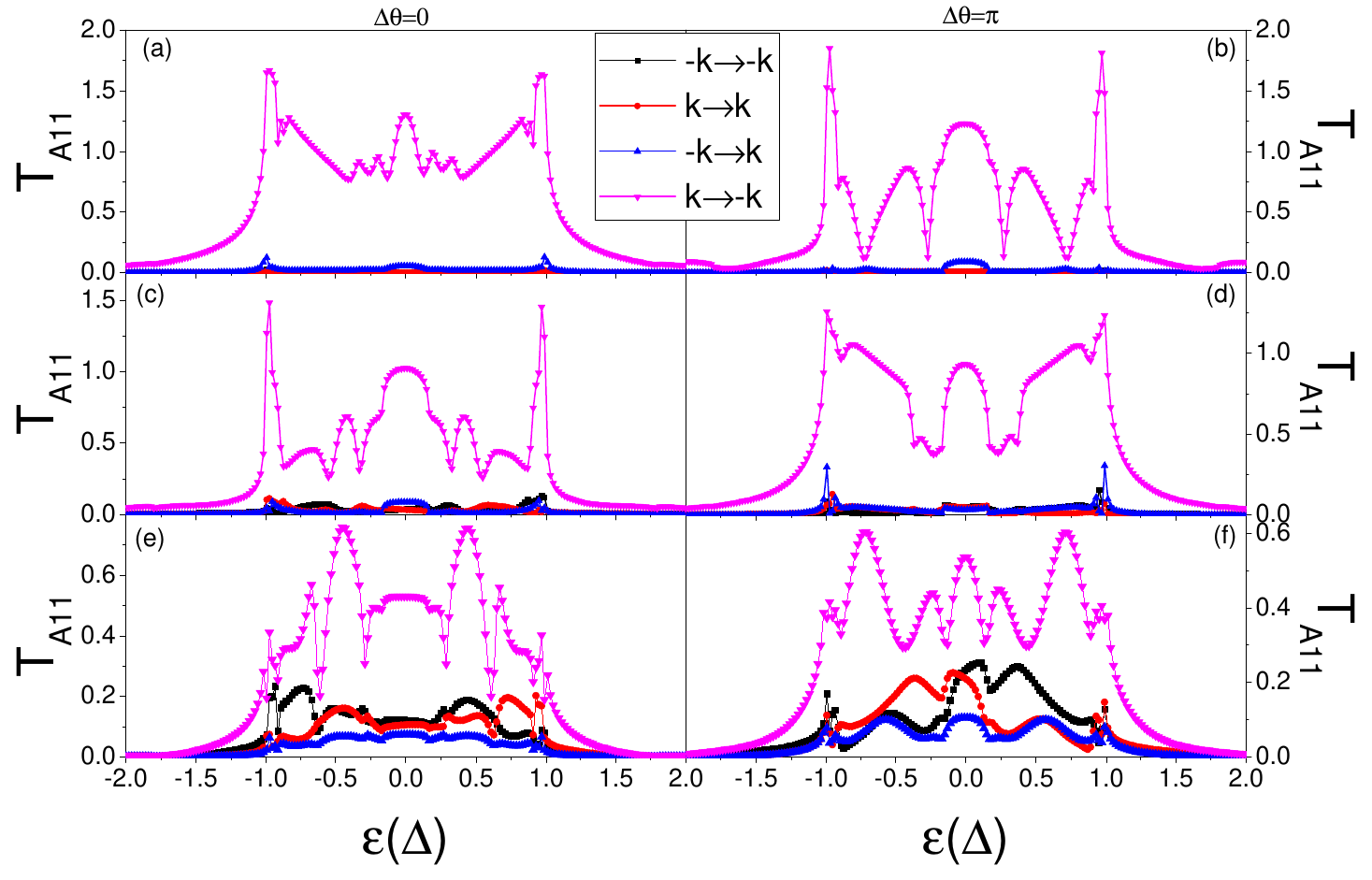}
\caption{(Color online) $T_{A11}$ vs the incident energy $\varepsilon$ in the four-terminal device for three different center region. The phase difference between the superconductor terminals are $\Delta\theta=0$ in (a) (c) and (e), and $\Delta\theta=\pi$ in (b) (d) and (f).  The structure of the center region are primitive graphene in (a) and (b),  Y-shaped Kekul\'{e} superlattices with $U_{A_3}=0$ in (c) and (d),  and  superlattices with $U_{A_3}=0.5t$ in (e) and (f). The length of the centre region is $L=50$ and the width is $W=80$. } \label{fig6}
\end{figure}

In order to investigate the variation of valley polarization in Andreev reflection within a three-terminal device featuring a Y-shaped Kekul\'{e} superlattice, both the incident energy $\varepsilon$ and on-site energy $E_{0}$ are simultaneously adjusted. As shown in Fig. \ref{fig5}, distinct boundaries can be observed for both $P_{A11}$ and $P_{A12}$, dividing the image into two regions with positive and negative values of valley polarization, respectively. In proximity to these boundaries, $P_{A11}$ and $P_{A12}$ can attain extreme values of $\pm1$. By manipulating the on-site energy $E_{0}$, it becomes possible to control the distance between the Dirac point and Fermi energy within the energy band. In Fig. \ref{fig2} (a), when the Fermi level intersects only with the first band, both the incident electron and the reflected hole possess a fixed valley degree of freedom, resulting in maximum valley polarization. By increasing $|E_{0}|$ to sufficiently large values, such as $E_{0}=-5\Delta$, due to intervalley scattering in the Y-shaped Kekul\'{e} superlattice, the reflected hole can have either $K$ or $-K$ as its valley degree of freedom when it flows into the graphene terminals. It is reasonable for a slight decrease in valley polarization for large $|E_{0}|$. In Fig. \ref{fig5}, it is evident that the sign of valley polarization primarily depends on the on-site energy $E_{0}$ in devices with Y-shaped Kekul\'{e} superlattices.

\begin{figure}[tbp]
\includegraphics[width=1.0\columnwidth,angle=0]{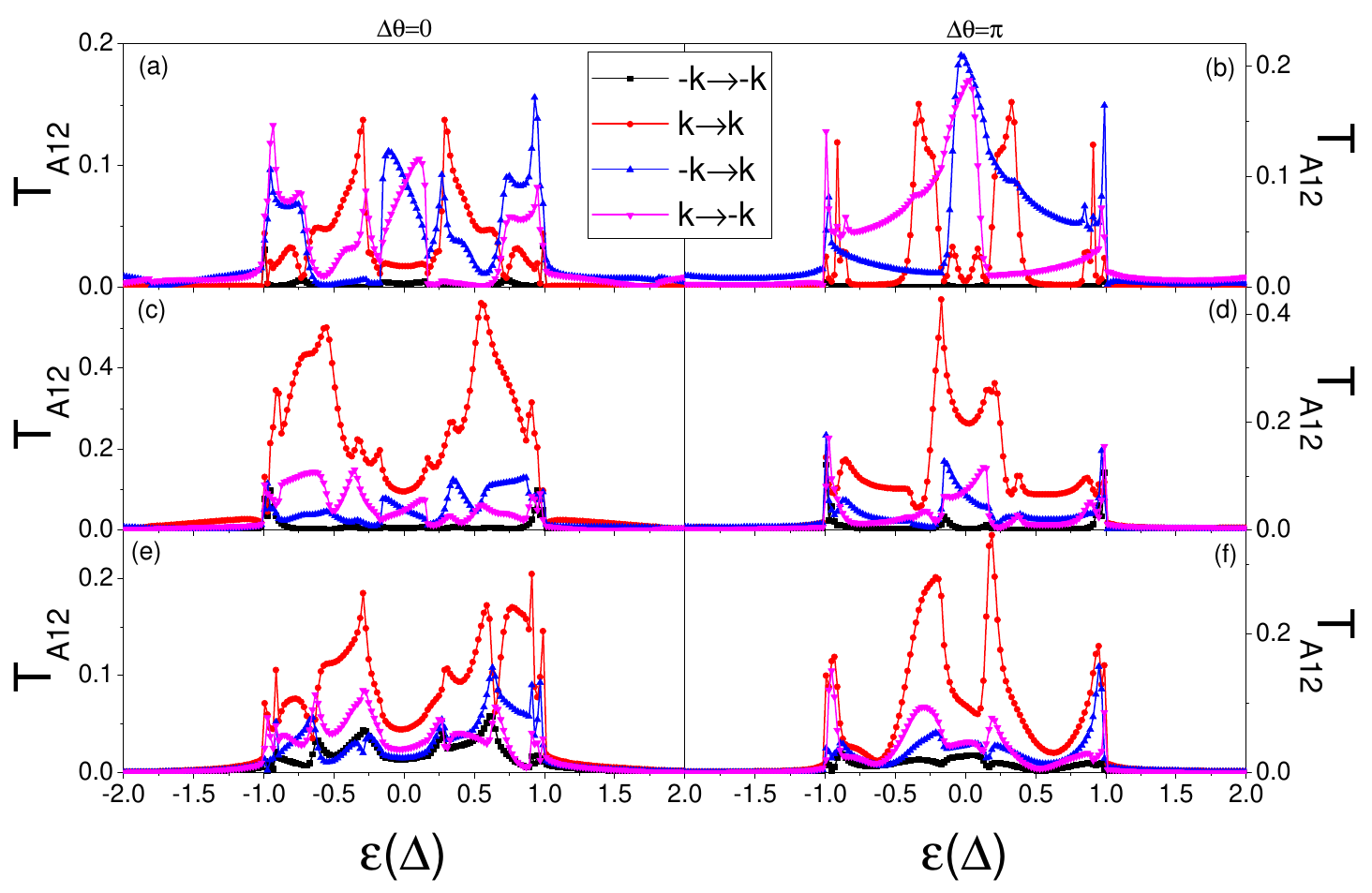}
\caption{(Color online) $T_{A12}$  vs the incident energy $\varepsilon$ in the four-terminal device for three different center region. The phase difference between the superconductor terminals  are $\Delta\theta=0$ in (a) (c) and (e), and $\Delta\theta=\pi$ in (b) (d) and (f).  The structure of the center region are primitive graphene in (a) and (b), Y-shaped Kekul\'{e} superlattices with $U_{A_3}=0$ in (c) and (d),  and  superlattices with $U_{A_3}=0.5t$ in (e) and (f). The length of the center region is $L=50$ and the width is $W=80$. } \label{fig7}
\end{figure}

In previous works\cite{Cheng,wangchao2020,zhang2022}, Andreev reflection can be affected by the phase difference between the two superconductor terminals in the four-terminal device. In the four-terminal device, the coefficients of Andreev reflection, namely $T_{A11}$ and $T_{A12}$, exhibit variations with incident energy under different superconductor phase differences $\Delta \theta$, as depicted in Fig. \ref{fig6}  and Fig. \ref{fig7} respectively. By comparing the coefficients at $\Delta \theta=0$ and $\Delta \theta=\pi$, it becomes evident that the superconductor phase difference significantly impacts Andreev reflection. Considering a primitive graphene structure for the central region, we calculate the coefficients of $T_{A11}$ presented in Fig. 6 (a) and (b). Regardless of the magnitude of the superconductor phase difference, $T^{K,-K}_{A11}$ consistently surpasses other coefficients. Within the range of $0.5\Delta<|\varepsilon|<1.0\Delta$, there is a remarkable decrease in $T^{K,-K}_{A11}$ when adjusting from $\Delta \theta=0$ to $\Delta \theta=\pi$. By comparison, the adjustment of the superconductor phase difference has little effect on the intravalley reflection $T^{K,K}_{A11}$ and $T^{-K,-K}_{A11}$, whose values are always tiny.  Then, the central region undergoes a structural transformation into a Y-shaped Kekul\'{e} superlattice, as depicted in Figs. \ref{fig6} (c) and (d). It is observed that within the range of $0.5\Delta<|\varepsilon|<1.0\Delta$, adjusting the phase difference of the superconductors from $\Delta \theta=0$ to $\Delta \theta=\pi$ enhances $T^{K,-K}_{A11}$. However, upon introducing $U_{A_{3}}=0.5t$ into the superlattice, there is a decrease in the peak value of $T^{K,-K}_{A11}$ and its dependence on the superconductor phase difference becomes limited. Comparing Figs. \ref{fig6} (e) and (f), it can be observed that both coefficients for intravalley reflections, namely $T^{K,K}_{A11}$ and $T^{-K,-K}_{A11}$, are consistently lower than $T^{K,-K}_{A11}$. Nevertheless, when the superconductor phase difference is set at $\Delta \theta=\pi$, these coefficients no longer exhibit symmetry about $\varepsilon=0$.

Under the different superconductor phase difference $\Delta \theta$  the variation of $T_{A11}$  has been discussed in the four-terminal device. In Fig. \ref{fig7} it shows how the coefficients of $T_{A12}$ change with the incident energy $\varepsilon$ under different superconductor phase difference. When the central region consists of primitive graphene, $T_{A12}$ exhibits distinct characteristics compared to the three-terminal device.  Specifically, irrespective of the phase difference between the superconductors, the intravalley reflection $T^{K,K}_{A12}$ exhibits a greater intensity compared to that observed in the three-terminal device. Moreover, for $\Delta \theta=\pi$, there is an enhancement in the intervalley reflections $T^{K,-K}_{A12}$ and $T^{-K,K}_{A12}$. However, when we consider a Y-shaped Kekul\'{e} superlattice as the structure of the central region in a four-terminal device, we observe that intravalley reflection remains as the dominant process for crossed Andreev reflection, similar to its behavior in a three-terminal device. The curves depicting $T^{K,K}_{A12}$ can be seen in Fig. \ref{fig7} (c) and (d). Around $\varepsilon=0$, there exists a minimum value of $T^{K,K}_{A12}$ when $\Delta \theta=0$. Conversely, $T^{K,K}_{A12}$ exhibits a peak value around $\varepsilon=0$ when $\Delta \theta=\pi$. It is evident that the strength of the crossed Andreev reflection coefficients can be controlled by adjusting the phase difference in the superconductors. Upon introducing $U_{A_{3}}=0.5t$ into the superlattice, the impact of the superconductor phase difference on $T_{A12}$ diminishes. The curves depicted in Fig. 7 (e) resemble those in Fig. 7 (f). From Figs. 6 and 7, it can be observed that the phase difference between two superconductor terminals influences Andreev reflection; however, this influence weakens due to broken chiral symmetry.

\begin{figure}[tbp]
\includegraphics[width=1.0\columnwidth,angle=0]{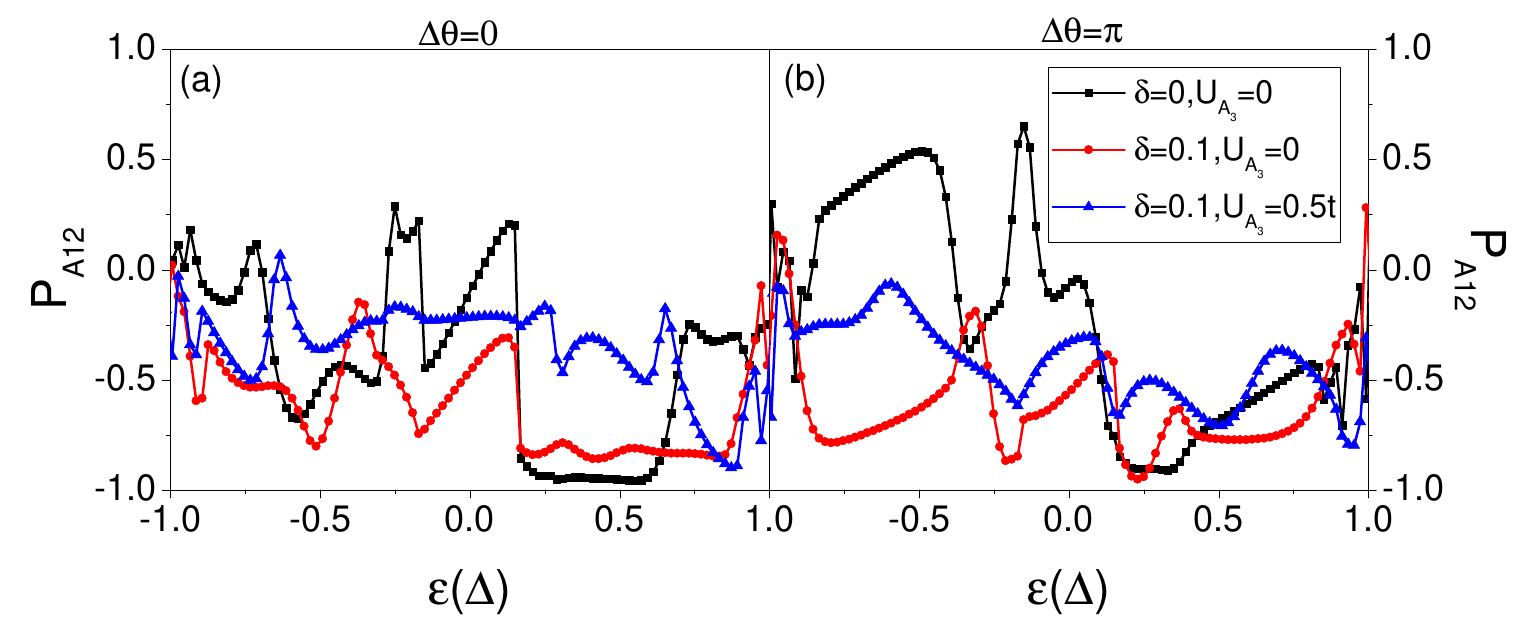}
\caption{(Color online) In the four-terminal device $P_{A12}$ vs the incident energy $\varepsilon$ for three different center regions. The difference of the superconductor phase are (a) $\Delta\theta=0$  , and (b) $\Delta\theta=\pi$, respectively.  The length of the centre region is $L=50$ and the width is $W=80$. } \label{fig8}
\end{figure}

The curves in Fig. \ref{fig8} illustrate the valley polarization of crossed Andreev reflection in a four-terminal device for two different phase differences between the superconductors, namely $\Delta \theta=0$ and $\Delta \theta=\pi$. In Fig. \ref{fig8} (a), except for the black curve, the values of $P_{A12}$ are consistently negative within the range of $|\varepsilon|<\Delta$. For the black curve, as $\varepsilon$ varies from $0.2\Delta$ to $0.6\Delta$, the valley polarization exhibits a plateau-like behavior approaching -1, indicating that this corresponds to a primitive graphene structure in the central region. However, when transitioning to a Y-shaped Kekul\'{e} superlattice structure in the central region, there is an increase in plateau value around -0.9 for the red curve. After incorporating $U_{A_{3}}=0.5t$ into the superlattice, for $-0.5\Delta<\varepsilon<0.5\Delta$, the value of $P_{A12}$ increases up to -0.5 as depicted by the blue curve. By setting the phase difference between the superconductors as $\Delta \theta=\pi$, Fig. \ref{fig8} (b) illustrates that the black curve exhibits a positive value within the range of $-\Delta<\varepsilon<0$ and transitions to a negative value for $0<\varepsilon<\Delta$. Notably, there is a plateau with a positive value around 0.5 observed for $-0.9\Delta<\varepsilon<-0.5\Delta$. Interestingly, within this same range, the red curve displays a plateau with a negative value around $-0.5$, indicating an inversion of valley polarization due to changes in the structure of the central region.

\section{Conclusions}
 In this study, we calculate the coefficients of intervalley and intravalley Andreev reflection in a multi-terminal device that we propose. The central region of the device consists of primitive graphene and a Y-shaped Kekul\'{e} graphene superlattice. From the energy band structure of the Y-shaped Kekul\'{e} graphene superlattice, it is evident that the two valleys are coupled together, resulting in enhanced intervalley scattering. Consequently, it is possible for both electrons and holes involved in Andreev reflection to originate from the same valley. In the three-terminal device, the coefficient of intravalley reflection $T^{K,K}_{A12}$ is greater than that of intervalley reflection when employing a Y-shaped Kekul\'{e} graphene superlattice as the structure in the center region. The valley polarization of Andreev reflection can be controlled by adjusting both the structure and on-site energy. In a four-terminal device, the valley-dependent transport of Andreev reflection can be influenced by manipulating the phase difference between superconductor terminals. To summarize, it is possible to manipulate the degree of freedom associated with valleys during Andreev reflection, providing novel ideas for designing and developing electronic components.

\setlength{\parindent}{0pt}
\section*{ACKNOWLEDGMENTS}
This work was supported by the National Natural Science Foundation of China (Grant No. 12274305 and No. 11874139), the Natural Science Foundation of Hebei (Grant No. A2022106001 and No. 2019205190).\\

%\begin{references}

%\end{references}

\end{document}